\documentclass[iop]{emulateapj}
\usepackage{natbib}

\shorttitle{Flows and Oscillation Modes}
\shortauthors{Baldner and Schou}

\begin{document}

\title{Effects of Asymmetric Flows in Solar Convection on Oscillation Modes}
\author{Charles S. Baldner and Jesper Schou}
\affil{W.W. Hansen Experimental Physics Laboratory, Stanford University, Stanford, CA 94305-4085, USA}
\email{baldner@stanford.edu}

\begin{abstract}
Many helioseismic measurements suffer from substantial systematic errors. A particularly 
frustrating one is that time-distance measurements suffer from a large center to limb 
effect which looks very similar to the finite light travel time, except that the magnitude 
depends on the observable used and can have the opposite sign \citep{Zhaoetal12}. This has 
frustrated attempts to determine the deep meridional flow in the solar convection zone, 
with \citet{Zhaoetal12} applying an ad hoc correction with little physical basis to 
correct the data. In this letter we propose that part of this effect can be explained by 
the highly asymmetrical nature of the solar granulation which results in what appears to 
the oscillation modes as a net radial flow, thereby imparting a phase shift on the modes 
as a function of observing height and thus heliocentric angle.
\end{abstract}

\section{Introduction}
The meridional flow in the solar convection zone plays a key role
in many solar dynamo models and an accurate measurement of the
flow with depth and latitude would thus be invaluable for
constraining solar dynamo models. There have been numerous 
attempts to obtain estimates using a variety of techniques, including  
time-distance \citep[in the following Z12]{Giles00,Zhaoetal12}, 
ring diagrams \citep{SchouBogart98,Haberetal02}, 
normal modes \citep[in the following W12]{Wetal12}, 
and supergranulation studies \citep{Hathaway2011}. 
While these studies have given reasonable numbers
near the surface, they have suffered from large and unexplained
systematic errors, preventing us from obtaining reliable numbers
throughout the convection zone.

A meridional counter-cell at high latitude was first found by 
\citet{Haberetal02}; subsequent work \citep{GHetal06,Zaatrietal06} 
found this to be a periodic phenomena tightly correlated with the 
solar inclination angle ($B_0$). \citet{Zaatrietal06}, concluding 
that the counter-cells were likely spurious, applied a correction 
to remove them. \citet{BraunBirch08} found that North--South 
travel times differed depending on the heliocentric longitude 
at which the measurement was performed. Most recently, Z12 used 
data from the {\it Helioseismic and Magnetic Imager\/} (HMI) to measure 
East--West travel time shifts along the equator and North--South 
travel time shifts along the central meridian in four different 
observables: continuum intensity, line core intensity, line depth 
(continuum minus line core), and Doppler velocity. They found 
large E--W travel time shifts along the equator, and found that they were 
quite different in different observables.

Although Z12 did not provide an explanation for the source of the 
error, they treated it as a heliocentric angle dependent phase 
or time shift. Using this assumption, they used the E--W travel 
time anomalies to correct the N--S travel time shifts --- this  
brought the four different observables into good agreement, 
which was encouraging. Similarly, it was noted by W12 that a radially 
varying phase of the eigenfunctions coupled with the variation in the 
height of formation with changing heliocentric angle might be an 
explanation, but again no source of such a phase variation was 
identified.

It is evident that helioseismic observations should suffer from a 
phase error as a function of heliocentric angle --- as \citet{DH09} 
pointed out, the light travel time for an observation at disk 
center is different than for an observation near the limb by 
roughly 2s. When analyzing travel time residuals, however, they 
found the correct magnitude but opposite sign. They conlcuded that 
the measured travel times suffer from a systematic effect with twice 
the magnitude and opposite sign as the expected light travel time 
effect. Similar numbers were found by 
\citet{SchouWoodardAAS12}, who found that a travel time error of 
2--3s at a heliocentric angle of 60$^\circ$ could explain their 
results.

In general, standing acoustic waves should 
have a constant phase with height. Many effects that are known to be 
poorly modeled or neglected entirely  do not change this. What is 
required to add a phase shift 
with height is an asymmetric effect, e.g. an effect which knows 
whether a wave is traveling upwards or downwards. In the Sun, of 
course, the modes we observe are not purely standing, and mode 
damping due to, for example, non-adiabatic effects will have an effect. In this 
work we are considering low frequencies, however, so we neglect this.

Here we suggest that a phase variation arises from the large asymmetry 
between the upflows and downflows in the convection near the solar 
surface. In particular, the broad upflows and narrow downflows give 
rise to net vertical flows when horizontally averaged over length 
scales much smaller than the acoustic modes. As was shown by \citet{GH2010}
in the context of meridional flows, 
a flow introduces a phase shift in acoustic modes --- one would 
expect the vertical convective flows in the Sun to introduce a phase 
shift with height in the solar atmosphere.

In the following discussion, we extract complex eigenfunctions from 
a detailed numerical simulation of convection to compute the phase 
delay as a function of height. We attempt to explain this phase shift 
as being due to the spatially averaged vertical flows by computing 
phases shifts to theoretical eigenfunctions with an imposed vertical 
flow and comparing these to the numerical eigenfunctions from the 
simulation. We use the phase shifts from the simulation data to 
predict systematic effects in travel-time measurements as a function 
of distance from disk center, and compare these effects to the 
observed discrepancies in solar data. Finally, we discuss some of the 
shortcomings of our models, how they might be improved, and discuss 
other observable consequences.

\begin{figure}
\plotone{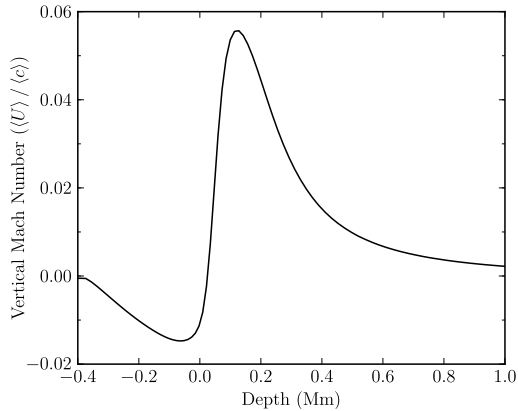}
\caption{Mach number as a function of depth from a model by 
\citet{Steinetal09}. 
The area average of the flow divided by the area average 
of the sound speed is used.
To exclude the buffer zone used in the model the outer 5 grid points
were excluded. The Mach number below 1Mm is negligible.
\label{mach96}}
\end{figure}

\section{Models}
We use a non-magnetic simulation of solar convection 
\citep{Steinetal09}.\footnote[1]{The model can be found at 
http://sha.stanford.edu/stein\_sim/ }
Convection in the outer layers of the solar envelope and atmosphere are 
simulated in a small Cartesian box.  We show 
the horizontally averaged vertical Mach number in Figure \ref{mach96}. 
In this work we assume that the modes `see' only this horizontal 
average. For low degree modes, this is likely to be an adequate 
approximation for our purposes.

\subsection{Simulation Results}
\label{sec:simresults}
\begin{figure}
\plotone{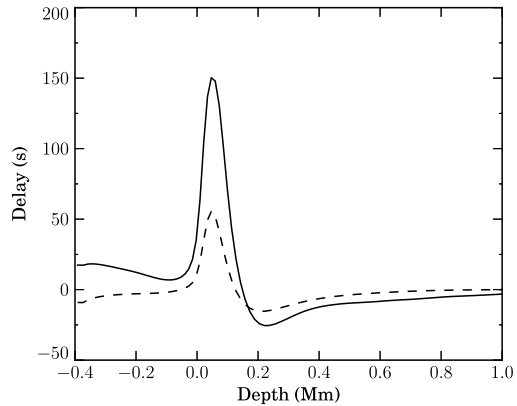}
\caption{Time delay as a function of depth from the simulation.
Solid line is for lowest $n$ at 1142$\mu$Hz, dashed for the second lowest $n$
at 1761$\mu$Hz.
\label{plotnum1}}
\end{figure}

The phase shifts of the eigenfunctions can be estimated by
extracting them from a convection simulation, as previously done
\citep{SteinNordlund2001}, but this time including the imaginary component.
This is easily done by
horizontally averaging the vertical velocity (to isolate radial modes)
from such a simulation and taking a temporal Fourier transform at each depth. 
The phase shifts can be converted to time delays by dividing by 
the angular frequency $\omega$.
Results of such an analysis
are shown in Figure \ref{plotnum1} for the two lowest frequency modes 
(those at higher $n$ are less well defined).

\subsection{Explaining the Simulation Results: Theoretical Eigenfunctions}
\begin{figure}
\plotone{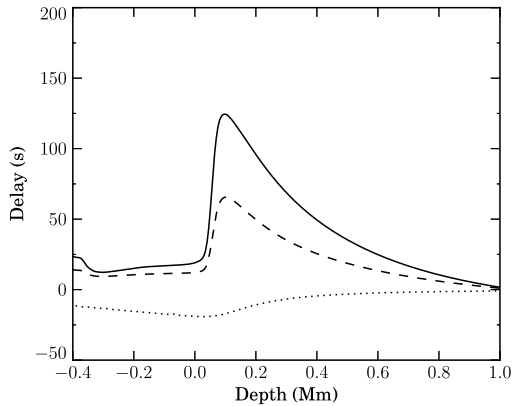}
\caption{Predicted effect of vertical velocity on mode eigenfunctions. The 
dotted line shows the result of integrating Equation \ref{pipe}, multiplied 
by a factor of 10 to that it can be seen on this scale. The solid and dashed 
lines show the predictions from a more detailed calculation, at 1142 $\mu$Hz 
and 1761 $\mu$Hz, respectively.
\label{eigencalc}}
\end{figure}

What is the cause of the phase shift we see in the standing waves in the 
simulation box? To get a crude estimate of the effect of the vertical flows, 
we start by considering the wind-in-a-pipe model of \citet{GH2010}, Section 3. 
Manipulating their equations and extending their Equation (9) to allow all variables 
to depend on position shows that the time shift introduced by a flow is given 
by
\begin{equation}
\Delta \tau (x) = \int_{x1}^{x2} \frac{U(x)}{c(x)^2} dx = \int_{x1}^{x2} \frac{M(x)}{c(x)}dx,
\label{pipe}
\end{equation}
where $x$ is the distance from the end of the pipe. The results of this 
calculation are shown in Figure \ref{eigencalc}. It will be noted that, while 
some features of the numerical eigenfunctions are qualitatively reproduced 
(i.e. a negative slope in the atmosphere and a steeper positive slope just 
below the photosphere), there is a quantitative disagreement of more than an 
order of magnitude. This is not surprising, however, as this toy model 
assumes that the background model varies slowly compared to the wavelength, 
and that is manifestly not the case here.

A somewhat more sophisticated approach is to compute the oscillation 
eigenfunctions for a stellar envelope with a specified vertical flow 
$U(r)$. To do this, we perturb the 
equations of continuity, motion, and energy in the usual way 
\citep[following, e.g., ][]{Unnobook}, but with the added velocity 
term in the base equations. For simplicity, we consider only radial 
modes in this work, but the generalization to non-radial modes is straightforward. 
When a vertical flow is present, the solutions become complex, and we can 
compute a phase delay as a function of height.

By including only a horizontally averaged vertical flow, we are of course 
involving a certain physical inconsistency --- that is, a horizontally 
invariant vertical flow is not consistent with a one dimensional background model, 
which requires a zero net mass flux. Because we are computing eigenfunctions 
in the linear perturbative regime, this is not necessarily an unreasonable 
inconsistency to accept, but it does require that we employ certain 
assumptions. As noted above, we assume here that the modes `see' the 
horizontal spatial average of the convective flows. One consequence of 
this assumption is that we assume that variations on convective length 
scales (horizontally) do not affect the acoustic modes. For radial modes 
this is reasonable. It also also assumes that the correlations between 
flows and thermodynamic quantities --- say, density --- do not have an 
effect on the phases of these modes.

Using the horizontally averaged thermodynamic quantities and vertical 
flow from the simulation box, we integrate the oscillation equations for 
radial modes with frequencies of 1142$\mu$Hz, and 1761$\mu$Hz, which may 
be directly compared with the eigenfunction phases shifts shown in Figure 
\ref{plotnum1}. The calculated eigenfunction phase 
shifts are shown in Figure \ref{eigencalc}. As can be seen, there is 
a general qualitative agreement between the two sets of phase delays. 
Most prominently, we see a strong positive phase shift just below the 
photosphere, though we calculate a somewhat smaller shift, and also 
find the peak to be much broader below the surface. In the atmosphere 
itself we find a fairly constant phase, or a slight phase lag with 
height. This is not entirely in agreement with the numerical eigenfunctions, 
but there is some uncertainty in the numerical eigenfunctions due to 
noise.  We consider
the agreement to be sufficiently good to conclude that the phase shifts
we observe in the convection simulations are due to the vertical
convective flows.

\subsection{Effects on observations}
The HMI instrument measures, among other things, continuum intensity, 
line core intensity, and line-of-sight velocity. These observables are 
produced at a range of heights in the solar atmosphere \citep{Flecketal11}, 
and can be represented as the convolution of some  
contribution function of height and the actual variation of the quantity 
in the solar atmosphere. A contribution function for the HMI line core 
intensity measurements can be found in \citet[][Figure 2]{Flecketal11}; 
the other observables have contribution functions that peak at different 
heights. The contribution function can be approximated by a Gaussian with 
a width of 250km. We will use this in the work that follows. A detailed 
study of the contribution function of the various HMI observables as a 
function of viewing angle is far beyond the scope of this letter.

To relate the change with height to the change with viewing angle 
we need to determine the relationship between these two quantities. Assuming 
that the atmosphere is isothermal and that the height of formation corresponds 
to the place where a certain column density of matter has been traversed from 
infinity, it can easily be shown that the change in height with angle is 
given by $\Delta h = - H \log(\cos(\theta))$, where $H\approx$120 km
is the density scale height and $\theta$ is the angle between vertical
and the line of sight. At $\theta=60^\circ$ this corresponds to about 80km. 

\begin{figure}
\plotone{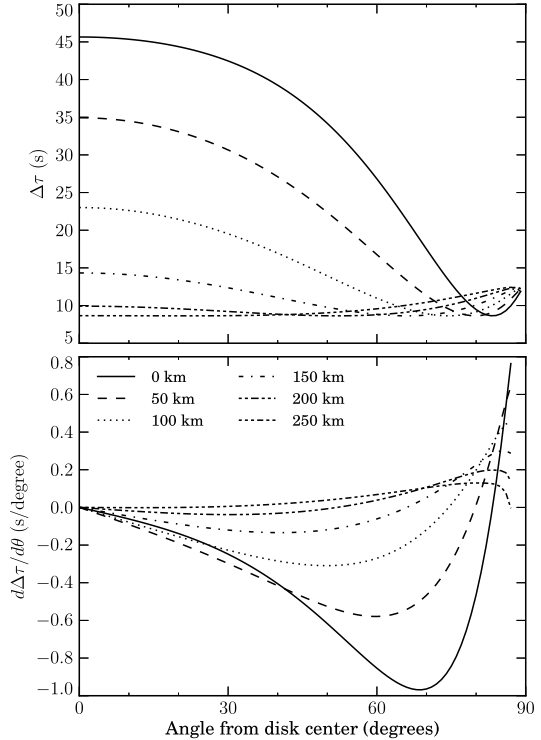}
\caption{Phase shifts as a function of viewing angle for different disk 
center formation heights, integrated over a Gaussian contribution function. 
The width of the Gaussian is 250km, chosen to match the contribution 
function from \citet{Flecketal11}. The top panel shows the phases at 
different viewing angles; the bottom panel shows the derivative of the 
phases with viewing angle.
\label{phaseshift}}
\end{figure}

In Figure \ref{phaseshift} we show the time delays integrated over 
a Gaussian contribution function as a function of viewing angle for 
various disk-center formation heights. We also show the derivatives 
with respect to angular distance from disk center of the time delays. 
These derivatives give the travel time differences one would measure in 
the limit of infinitely small apperture size, and can provide a good 
approximation for the apperture sizes used in Z12. For direct comparison 
with Z12, the values of our derivatives must be multiplied by the 
apperture size used in the measurement.

\section{Discussion}
We have explored the effects of vertical convective flows on solar acoustic 
modes. Can they explain the systematics that have been observed in solar data, 
for example the effect found in Z12? In that work, 
the authors showed East-West travel time differences across the the solar 
equator for different HMI observables (line-of-sight velocity, continuum 
intensity, line core intensity, and line depth). They found that all 
observables exhibited systematic effects along the equator that they 
believed to be spurious, but that each different observables had different 
signatures. Comparing their Figure 2 with the bottom panel of 
our Figure \ref{phaseshift}, we 
note some striking similarities. First, the continuum shows the largest 
anomaly --- as does the 0km height measurement we predict, and with the 
same sign (negative). The Doppler signal, with a peak in the contribution 
function at approximately 150km, shows a much smaller anomaly but still 
the same sign. We predict the same thing (note that our 150km signal changes 
sign, but only above 60$^\circ$ from disk center, which is further than 
the Z12 results go. Finally the line core intensity, with the highest 
formation height (250km), shows an anomaly with the opposite sign, and this is 
matched by our results. Furthermore, the Z12 anomalies show turnovers 
as the distance from disk center gets large, and we show the same.

The quantitative agreement is not as promising, however. In particular, 
the magnitude of the effect we predict, 
by multiplying the derivative of the phase shifts by the aperture 
size used in the measurements, is too small --- at most a few 
seconds, as opposed to the more than 10 seconds Z12 find in continuum 
intensity. Furthermore, while we do predict the turnover observed, we 
do not accurately predict where this should happen.

Several approximations may be identified, some of which are
easily amenable to improvement and others not.
First of all the use of the area weighted vertical velocity is
unlikely to be correct. In reality the propagation of waves in
a medium with small scale ($\ll$ wavelength) variations is a
complicated issue and the velocities would likely
have to be weighted in some other way. In particular, variations
in different quantities (significantly density and velocity)
are correlated, which will need to be taken into account.

Another problem is that the granulation is not static on the
timescale of the oscillations and that there are likely to
be interactions between the two. This should be well captured
by the simulations, but is probably difficult to model accurately --- at 
least for the present authors.

Finally we have ignored complex issues of radiative transfer.
In approximating the contribution function as a Gaussian, we have 
simplified the problem but done violence to the actual physics. For 
the present purposes we consider it sufficient, but the exact shape 
of the actual contribution functions (which are different both for 
different observables and for the same observable at different viewing 
angles) will have significant quantitative effects. It is not likely, 
however, that the qualitative results would be affected. At least 
this problem is well understood and has been addressed in detail for 
other purposes.

In addition to the phase shift with observing angle, there are
other possible observational consequences.
Perhaps most obviously there will be a phase shift between
different observables, such as continuum intensity and Doppler
shift, which might be misinterpreted as a propagation or non-adiabaticity
effect.

Another effect is that even for the same observable there should be a
phase difference between observation heights. This includes, e.g.,
Doppler shifts derived from different parts of the spectral line and
even observations in the middle of granulation versus intergranular
lanes.

In addition to the observational effects, the fact that eigenfunctions
can be determined with significant accuracy from numerical
simulations presents many opportunities, in particular ones
involving the ability to determine if the depth variation of
both the real and imaginary parts match models.
Potentially this could be used to test models of the
interaction of waves with granulation and non-adiabatic effects,
hopefully leading to a better understanding of the physics behind
such things as the surface terms currently being applied in an ad hoc
way in structure inversions.

\section{Conclusion}
We have shown that the effect of the vertical flows from convection in the 
outer solar convection zone and atmosphere do affect the quantities we 
observe in helioseismology. We have further shown that the systematic 
errors that have been observed can be qualitatively explained by this 
effect. We conclude that the ad hoc correction applied by Z12 is likely 
justified.

A full quantitative prediction of the effects of the vertical flows 
requires a more sophisticated effort than that employed in this work. 
In particular, the modeling of the structure of the atmosphere must be 
very accurate and a proper treatment of the actual measurements we take 
must be done. This work, while non-trivial, is certainly feasible, and 
would be useful in addressing a number of different outstanding 
problems in helioseismology.

\acknowledgments
We are grateful to Bob Stein for help with the simulations
and to Regner Trampedach, Douglas Gough, Phil Scherrer, and Thomas Straus for 
useful discussions. The present work was supported, in part, by NASA contracts
NAS5-02139 and NNH09CF93C. The simulations used in this work were performed 
by Robert F. Stein, Ake Nordlund, Dali Georgobiani and David Benson. Their work was 
supported by NASA grants NNG04GB92G and NAG 512450, NSF grants AST-0205500 
and AST-0605738, and by grants from the Danish Center for Scientific Computing. 


\end{document}